\documentclass[aip,reprint,10pt]{revtex4-1}

\usepackage{graphicx}
\usepackage{dcolumn}
\usepackage{bm}
\usepackage[utf8]{inputenc}
\usepackage[T1]{fontenc}
\usepackage{mathptmx}
\usepackage{times}
\usepackage{natbib}
\usepackage{amssymb}
\usepackage{amsmath}
\usepackage{amsfonts}
\usepackage{color}
\usepackage{epstopdf}
\newcommand{\be}{\begin{equation}}
\newcommand{\ee}{\end{equation}}
\newcommand{\ba}{\begin{eqnarray}}
\newcommand{\ea}{\end{eqnarray}}
\newcommand{\bd}{\begin{displaymath}}
\newcommand{\ed}{\end{displaymath}}
\newcommand{\nnb}{\nonumber \\}
\renewcommand{\vec}[1]{\mbox{\boldmath$#1$}}

\begin{document}

\title{Radiation dominated implosion with flat target}
\bigskip


\author{
	L\'aszl\'o P. CSERNAI\! $^{1,3}$,
M\'aria CSETE\!  $^2$,
Igor N. MISHUSTIN\!  $^{3,7}$, 
Anton MOTORNENKO\!   $^3$, \\
Istv\'an PAPP\! $^4$,
Leonid M. SATAROV\!  $^{3}$,
Horst ST\"OCKER\!  $^3$,
and
Norbert KRO\'O\! $^{5,6}$\\
}

\affiliation{%
\small{$^1$ Dept. of Physics and Technology, Univ. of Bergen, Norway}\\
\small{$^2$ Dept. of Optics and Quantum Electronics, Univ. of Szeged, Hungary}\\
\small{$^3$ Frankfurt Institute for Advanced Studies, Frankfurt/Main, Germany}\\
\small{$^4$ Dept. of Physics, Babes-Bolyai University, Cluj, Romania}\\
\small{$^5$ Hungarian Academy of Sciences, Budapest, Hungary}\\
\small{$^6$ Wigner Research Centre for Physics, Budapest, Hungary}\\
\small{$^7$ National Research Center "Kurchatov Institute" Moscow, Russia}
}%
\date{\today \ \ \ \  {\color{red} Vs 1.04 pwp}} 
\begin{abstract}
	Abstract: Inertial Confinement Fusion is a promising option to provide massive, clean, and affordable energy for humanity in the future. The present status of research and development is hindered by hydrodynamic instabilities occurring at the intense compression of the target fuel by energetic laser beams. A recent proposal by Csernai et al. \cite{CKP-LPB-18} combines advances in two fields: detonations in relativistic fluid dynamics and radiative energy deposition by plasmonic nano-shells. The initial compression of the target pellet can be eliminated or decreased, not to reach instabilities. A final and more energetic, short laser pulse can achieve rapid volume ignition, which should be as short as the penetration time of the light across the target. In the present study, we discuss a flat fuel target irradiated from both sides simultaneously. Here we propose an ignition energy with smaller compression, largely increased entropy and temperature increase, and instead of external indirect heating and huge energy loss, a maximized internal heating in the target with the help of recent advances in nano-technology. The reflectivity of the target can be made negligible, and the absorptivity can be increased by one or two orders of magnitude by plasmonic nano-shells embedded in the target fuel. Thus, higher ignition temperature and radiation dominated dynamics can be achieved. Here most of the interior will reach the ignition temperature simultaneously based on the results of relativistic fluid dynamics. This makes the development of any kind of instability impossible, which up to now prevented the complete ignition of the target. 
\end{abstract}
\maketitle

\section{Introduction}

Inertial Confinement Fusion (ICF) is an ongoing activity aiming
for ignition of small pellets of thermonuclear, deuterium-tritium (DT) 
fuel by high-power lasers. The main direction of activity aims for 
strong compression of the fuel, where the resulting adiabatic 
heating would ignite the fuel 
in a central hot spot. 
The pulse and the compression
should be large and strong enough to keep the compressed fuel 
together for sufficient time for
burning after ignition, 
due to the inertia
of the compressed pellet. This is the aim of both the direct 
drive and indirect drive experiments.

In the present paper we work out the earlier presented idea 
\cite{CKP-LPB-18,Patent}
how to achieve nearly simultaneous volume ignition in the 
majority of the target. The two fundamentally new ideas in this work
are the same, and to the benefit of the reader, to be able to
read the article continuously, we repeat these principle items also here.
However, now we also present the theoretical concept for a simplified,
easier accessible, and energetically more advantageous  
flat target configuration. This obviously reduces the possibility
to reach extreme compression, but our aim is to increase the target 
energy density, 
and for this purpose the flat target configuration is advantageous.

In the pellet the fusion reaction, 
$D+T \rightarrow n (14.1 {\rm MeV}) +\, ^4\!He (3.5 {\rm MeV})$,
takes place at a temperature of $kT \approx 10$ keV. The produced
$^4\!He$ (or $\alpha$) particles are then deposited in the hot
DT plasma and heat it further. This is the plasma self-heating
(or $\alpha$-heating). The compression wave penetrates into
the plasma with the speed of sound or with the speed of a
slightly faster
compression shock. 

The goal of these experiments is to generate and sustain a self-propagating 
burn wave, which reaches the whole interior of the pellet.  
The present most advanced ICF experiments are done at the NIF, 
with spherical, $4\pi$, irradiation. Here we propose  
experiments simpler and more accessible for laboratories,
than the spherical irradiation setups at NIF, (192 laser beams)
\cite{Landen2012,RLL2012}
or OMEGA, (80 laser beams)
\cite{Nora2015}.
In the NIF experiments, the goal is to achieve extreme compression,
and adiabatic heating. This results in extreme pressure, minimal 
entropy production, and the relatively smallest temperature increase.
The extreme pressure leads to rapid subsequent expansion and to
possible instabilities. 

In the present paper we {\bf demonstrate} the 
advantages of a method of short ignition pulse with plasmonic 
nano shells, for this configuration, with or without 
possible target pre-compression.
The basis of the pre-compression is that the rate of fusion 
reactions in homogeneous medium is proportional to
\be
n_1 n_2   < \sigma \vec{v} >,
\label{rate}
\ee
where $n_1$ and $n_2$  are densities of D and T nuclei,
$\sigma$ is the reaction cross section and  
$\vec{v}$ is the relative velocity. 
In the proposed rapid, linear ignition the relative velocity is not
the thermal velocity, but the higher relative velocity of the two opposite
beams!

The effect of higher density is also useful, 
as we will discuss it later.	


Below,
we suggest testing of the method with two laser beams 
pointing against each other
with simultaneous irradiation pulses from both sides
of the target. Such a configuration was tested recently
successfully
\cite{Bonasera2019},
however without two fundamentally new aspects, the nearly simultaneous
heating to ignition on a so called time-like hypersurface, with a final
short energetic laser pulse and embedded nano-antennas in the target
to modify light absorption to achieve nearly uniform heating in the
whole target volume.

We think that a realistic calculation may be made only by 
relativistic fluid dynamics (RFD).
These velocities are significantly less
than the speed of light, $c$, so why would we need relativistic fluid 
dynamics to describe these reactions? 
RFD must be used not only with
high velocities or high velocity gradients, but also for radiation dominated
processes i.e. at high temperatures, when the energy density and
the pressure are of the same order of magnitude and not dominated by the
rest mass of the matter. RFD has qualitatively different features, proven
theoretically and experimentally. In particular, 
detonation or burning fronts can be 
both space-like and time-like (i.e. simultaneous in space-time) 
\cite{C87}.
Also in radiation dominated processes, fluctuations of the burning front 
are smoothed out because radiation will transfer energy to volume elements
with smaller energy density, which are created by mechanical flow
fluctuations
\cite{ZR1966}.

Linear laser irradiation will increase the density of transparent target,
as pointed out in case of ultra-relativistic heavy ions 
\cite{GyCs1986}, and also indicated by ref. 
\cite{Bonasera2019}.  This mechanism is different from the ablator 
technique used in spherical geometry, and suppresses RT instability.
Furthermore, the linear geometry leads to a non-isotropic momentum 
distribution where the accelerated ions are primarily moving in the 
beam direction like in colliding beam accelerators. 
Thus we avoid using the term temperature, and the term heating
referes to the increase of the typical energy density but not 
to a thermalized isotropic momentum distribution, therefore 
we can use eq. (\ref{rate}) for the reaction rate.

\section{Considerations for the target}

In the following we perform semi-analytic calculations
disregarding the target's compression. 
Here we consider a simple target configuration, without ablator 
using the early examples in refs.
\cite{C87,CS14,Patent}.
These estimates were based on analytic, scaling solution of relativistic
radiation dominated fluid dynamics.
For simplicity we consider flat solid targets to get 
quantitative estimates. We choose targets of
$h=$ 0.1-0.2 mm thickness, 
with  deuterium-tritium (DT) ice and polylactic acid (PLA) or
cyclic olefin copolymers (COC) as target materials.
These latter two target materials serve to test the 
two fundamentally new ideas of this work

These targets have relatively small absorptivity. 
To absorb the whole energy of the incoming laser light on 
$\sim$ 0.1 mm length, 
we need an absorptivity of the order of 
$\alpha_K \approx 10^2 $ cm$^{-1}$. 
This is typical 
absorptivity of DT fuel for soft X-ray radiation of 1 nm wavelength,
see Fig. 2 of ref. 
\cite{Hu2014}.
For softer radiation one gets larger absorptivity, and the radiation
can be absorbed in the outer layers of the pellet.

As mentioned above we assume two counter propagating laser beams 
aiming at the target from both sides simultaneously.

\section{Simplified model for flat target}

Let us
consider a flat piece of matter, which is sufficiently
transparent for radiation.  The absorptivity of the  target matter is 
considered to be constant, such that the total energy of
the incoming light is absorbed fully when the light reaches the
opposite edge of the flat target. 
 This matter undergoes an exothermic
reaction if its energy density that can be characterized
by a temperature parameter exceeds $T_c$. In a colliding beam 
configuration the collision rate exceeds the one in thermalized matter.

\begin{figure}[h]  
\resizebox{0.9\columnwidth}{!}
{\includegraphics{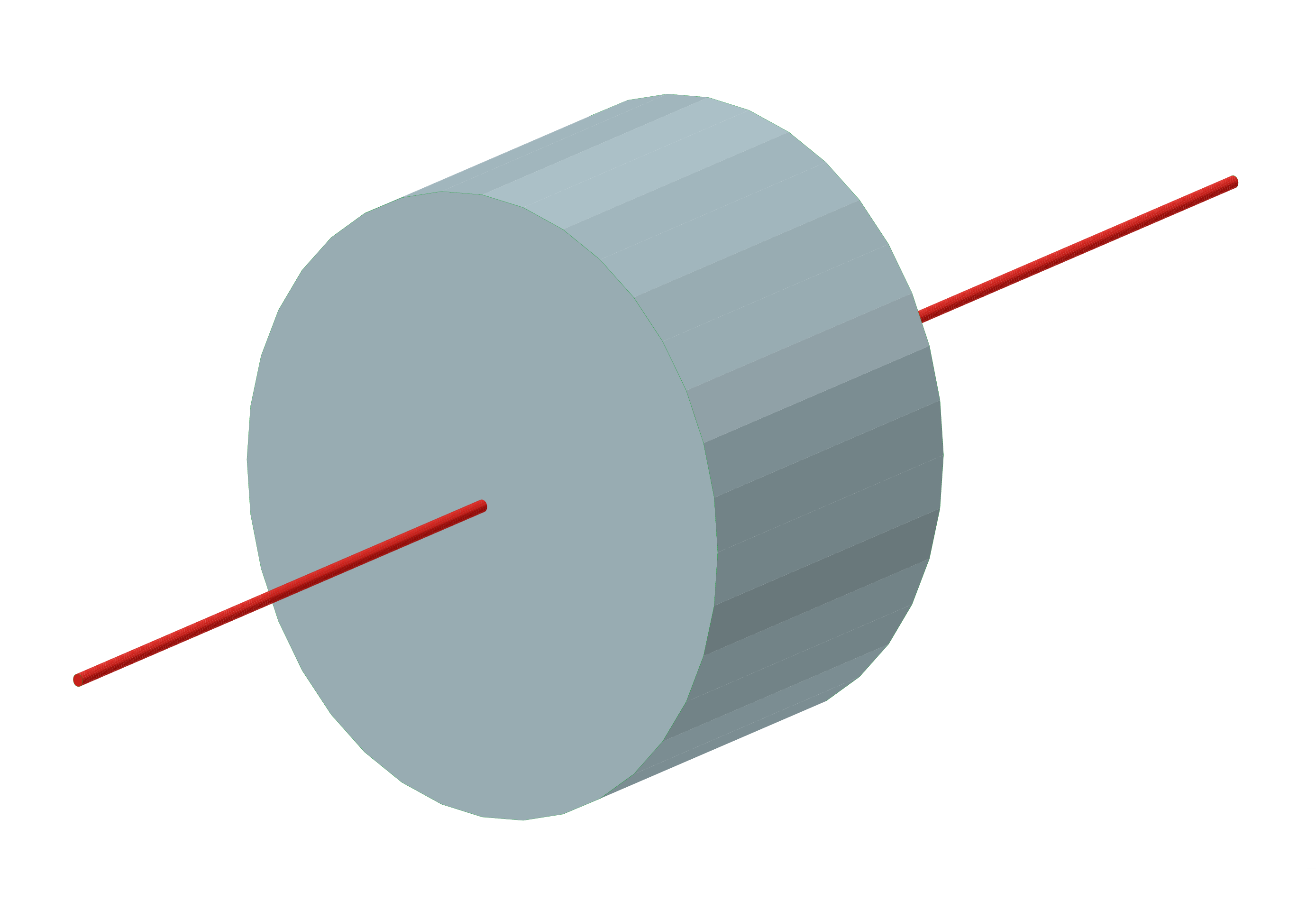}}
\caption{
(color online) 
Schematic view of 
the cylindrical, flat target of radius, $R$, and thickness, $h$, 
that is compact to minimize the surface effects.
The irradiation is performed along the $x-$axis from both sides towards the 
target. The laser beam should be uniform in the transverse direction, 
hitting the whole face of the coin shaped target.
}
\label{F-1}
\end{figure} 

The target matter is irradiated by two laser beams simultaneously
from the two opposite sides.
For simplicity of the presentation,
we are neglecting the matter expansion, so that the 
target thickness $h$ is taken to be constant. Experiment 
\cite{Bonasera2019} indicates that this is an acceptable 
assumption.

The laser beam profile is assumed to be uniform over the surface
of the target. This requires adequate optical properties, which
may be critical for short, 
(fs or ps)
pulses of high pulse energy and for a
$\sim$0.01 mm$^2$ 
target surface. For the lower energy test experiments and 
material technology studies, a\\
$\sim$1 mm$^2$ flat target 
surface size is sufficient. 

Below we consider a coin-shaped target with $h = D = 2R$, where
\be
V = 2\pi R^3, \ \ \ \ \
R = \sqrt[3]{V / (2\pi) }, \ \ \ \ \
h = \sqrt[3]{ 4V / \pi}.
\ee
 The laser irradiation comes from the 
$\pm x$ directions while the side surface of the coin allows
the laser light and particles to escape. This can be minimized by an 
external cover on the outside cylindrical 
edges made of a well reflecting material 
(like depleted Uranium-238) or with a metal of high melting 
point (like Tungsten). 

In this ICF configuration the energy deposition will depend
on the $x$ coordinate.
The simultaneous ignition can be achieved if 
the nano-shell concentration at the middle layer of the DT target 
increases faster than in the spherical configuration.
We consider two options:\\
{\bf (A)} Verification of the {\bf method of short (ps) ignition 
with plasmonic nano shells}, with a full volume transition
or melting of a non-explosive target material.\\
{\bf (B)} Simultaneous fusion with smaller amount of 
DT-ice target.

In both cases, the available laser pulse energy determines the 
possible outcome and the target size, which then provides us 
with the required ignition pulse length.
\medskip

{\bf (A)} Let us first consider option A. 
To test the basic principle of the method, we can  study a phase 
transition of a transparent material the cyclic olefin copolymers, 
(COC) or the polylactic acid (PLA).  We need transparent target 
for the penetration of the laser light. Let us take for example, 
the PLA. Its density is 1.21-1.43 g/cm$^3$, 
melting point 150-160 $^\circ$C, melting enthalpy 
28-38 J/g. 
Note that a 30 mJ 
laser pulse could melt 
1.07 mg of PLA.  
This amount of PLA has 
$V\sim$ 0.82 mm$^3$ volume. 
For a laser beam with 
0.81 mm$^2$ cross section 
the target thickness should be 
$h=D=$ 1.02 mm 
to achieve melting of the considered target with one single pulse.

In this case the penetration time of the 
laser pulse through the target is 
$t_x =$ 5.1 ps. 
considering a refractive index of 1.5.
To achieve homogeneous melting 
the target should be doped with resonant golden nano-shells, 
with enhanced concentration in the middle. Our calculation shows
(see below) that light absorption should be 
less on the flat target surface and more in the middle. 
 By varying the pulse length 
while keeping the pulse energy constant, one can see 
earlier melting at the outside edges (for shorter pulse length)  
or at the center (for longer pulse length).

For example the Ti-Sapphire HIDRA laser with 
~30mJ pulse energy at Wigner RCP
\cite{Wigner},
is adequate for such experimental verification 
of the method of short (ps) ignition 
with plasmonic nano shells. Similarly the PHELIX laser of GSI/FAIR
\cite{GSI}
is also adequate. In this case one has even a higher, (180 J), 
pulse energy for short, (ps), pulses.
\medskip

{\bf (B)} Estimating the pulse energy needed to ignite the DT fuel target 
is more complicated because of the spatial configuration of the target.  
We can make a rough estimate based on the results of the NIF 
experiments. Here 54 kJ fusion energy was detected 
\cite{NIF2018},
which is about twice as much as the invested ignition energy, 
for a DT target mass of about 1mg. 
The exact masses of the DT target varied in these 
experiments as well as the target structure. 
Detailed estimates of parameters are given in 
{\bf Appendix A}. 
We estimate an ignition energy of 
	about $Q/m \approx$ 27 kJ/(0.13 mg), 
\cite{NIF2018},
so that {\bf a laser pulse energy of 
Q=100J}, 
could ignite about 
$m = 0.5 \mu$g 
of DT fuel.  
For an uncompressed fuel density of 
$\rho =$ 0.225 g/cm$^3$ 
this would give a target volume of
 $V = 0.001-0.002 $mm$^3$. 
This is about 500 times 
smaller than the volume discussed above in case (A). 
The corresponding target diameter and thickness are 
$D=h=0.14$ mm. 
Assuming a refractive index of 1.13, 
the speed of light in the target is 
$c_{DT} = 0.27 \mu$m/fs, 
thus, the length of the ignition pulse, 
crossing the target, should thus be about 
$\Delta t = 0.5 $ ps. 

With highly intensive laser pulses we achieve a non-thermal situation
where along the lines of Laser Wake Field Acceleration (LWFA), ions are
compressed and accelerated in non-linear Wake waves or Wake bubbles,
up to GeV energies. The momentum distribution of ions is not thermal 
and not isotropic in this case but point in the $x$ -direction
against each other like in colliding accelerator beams.

The temporal shape of the laser pulses (with duration 
$t_x \sim $ 5.1 ps and 0.5 ps 
for cases A and B, respectively), is 
assumed to be rectangular in our simulations. 
The actual time profile of short pulse 
high power lasers is rather a Gaussian. This can also be modeled and
will not change the basic conclusions. 

\begin{figure}[h]  
\resizebox{1.0\columnwidth}{!}
{\includegraphics{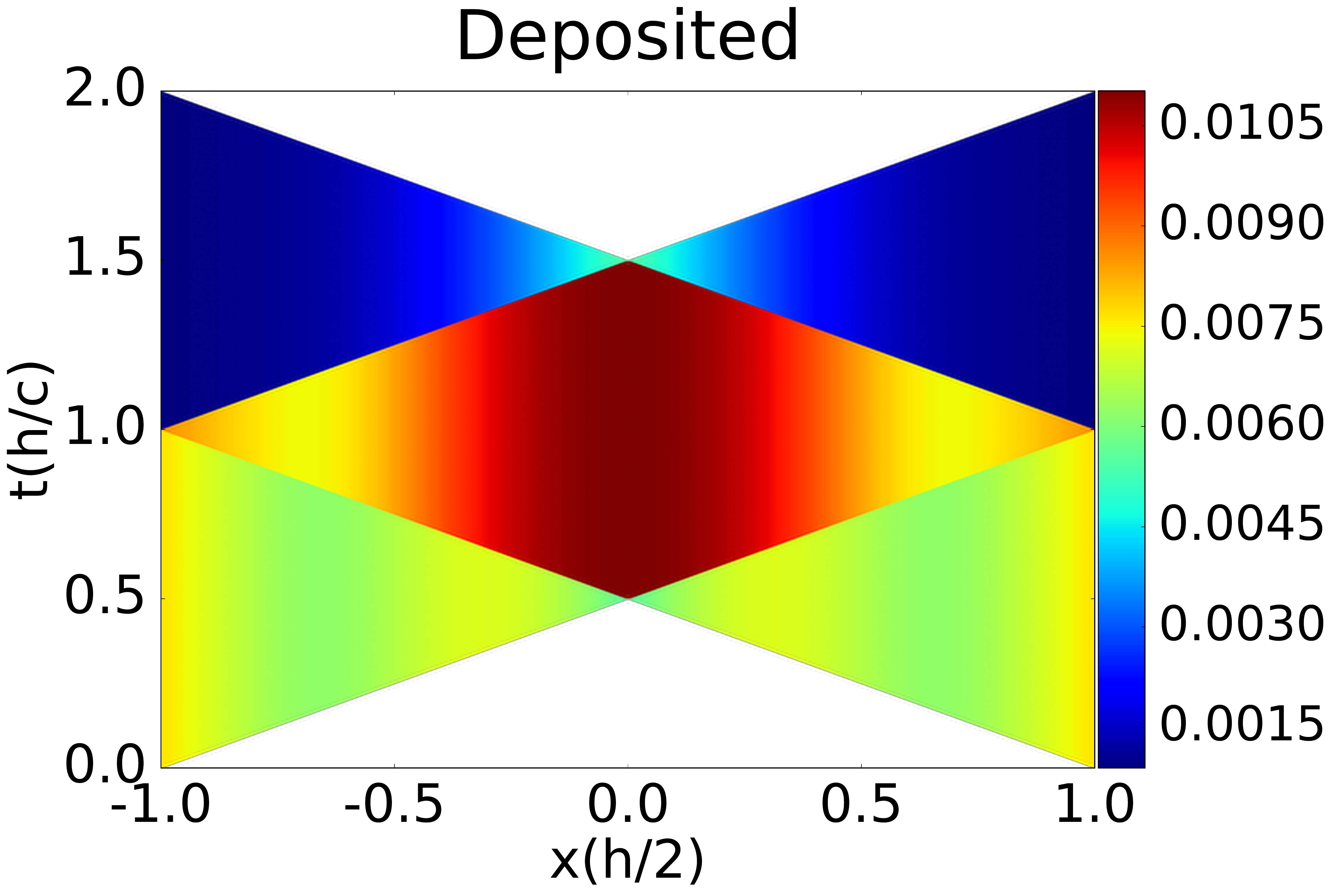}}
\caption{
(color online) 
Deposited energy per unit time in the space-time plane
across the depth, $h$, of the flat target.  The time is measured 
in units of $(h/c)$, where $c$ is the speed of light in 
the material of the target. The irradiation lasts for a 
period of $\Delta t = h/c$ the time needed to cross the 
target. The irradiated energy during this time period is 
$Q$ from one side, so it is $2Q$ from both sides together.
The color strip shows the deposited energy per unit time 
and unit cross section (a.u.).  Note that in this calculation
the absorptivity is $\alpha_K \not=\ \  $const. For more details 
please see Appendix B. 
}
\label{F-Depos}
\end{figure} 

Our choice for the
pulse energy and pulse length are within the starting plans 
of ELI-NP (1/min, 92J, 22fs, 810nm) and within the future extension 
plans of ELI-ALPS. This latter institute is in the process of 
installation of smaller energy and shorter pulse length laser 
(10Hz, 2.5J, < 17fs, 810nm). This makes it possible to study 
a ~3 times smaller target, with a 3 times shorter 
ignition pulse. This is possible with the planned shorter, fs, 
pulse length of this laser.

Below we calculate the energy density distribution, characterized by
a parameter
$T(t,x)$, at the initial stages of cylindrical target evolution, 
as a function of time, $t$, and the distance from
the center plane of the target $x$.  
Our calculation is made in two steps: 

(i) At the first step we calculate how much energy can reach 
a given point, $x$, in the target. 
The radiation starts at time $t = 0$, from the outer surface.
The eventual, "Low foot" type pre-compression is not included 
in the
present dynamical calculation, but we can start 
the simulation from a pre-compressed target state also.
See Fig. \ref{F-Depos}.

\begin{figure}[h]  
\resizebox{1.0\columnwidth}{!}
{\includegraphics{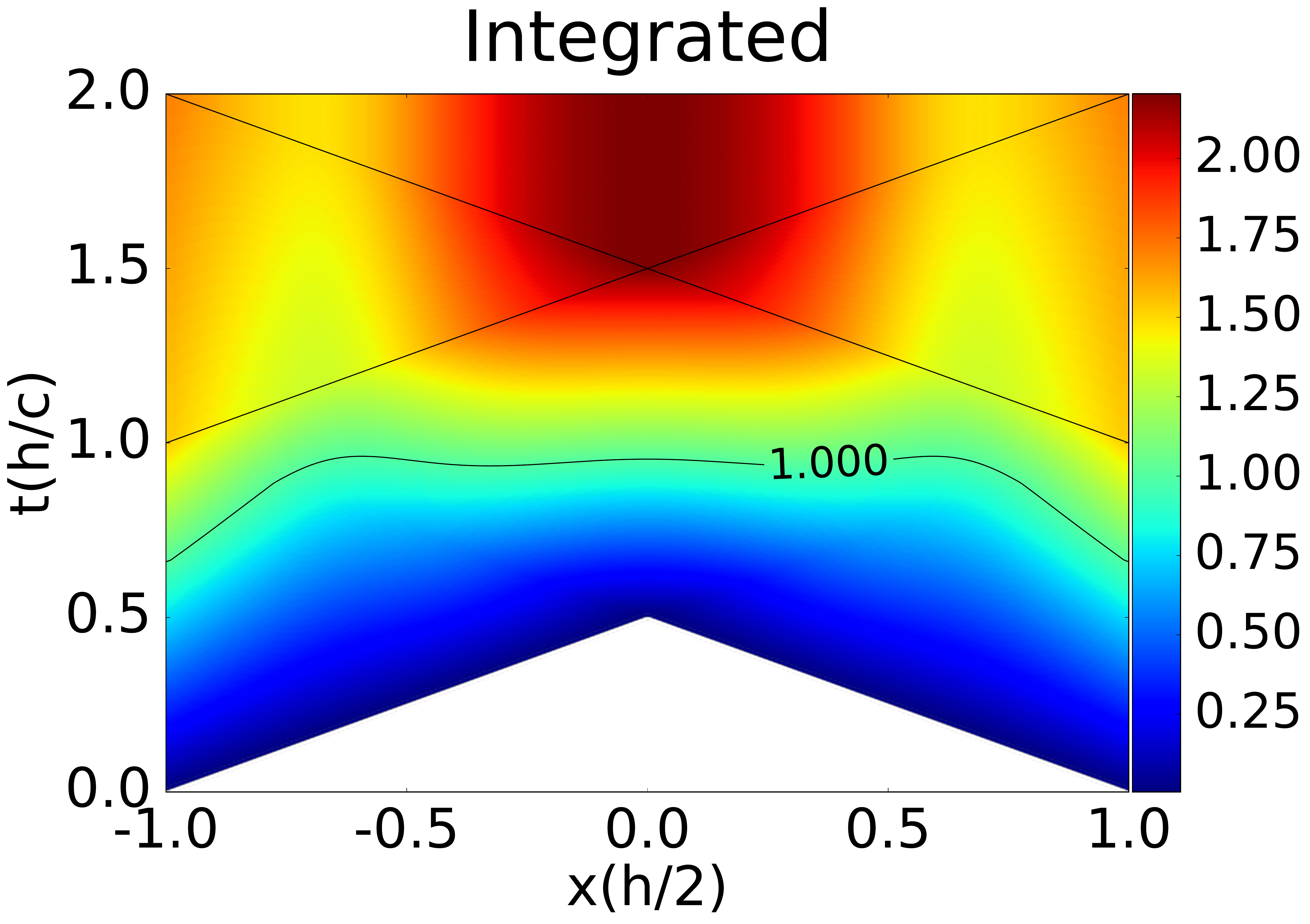}}
\caption{
(color online) 
The energy density $T(t,x)$ achieved up to a given time 
in the flat target. The color strip
indicates the energy density, $T$, in units of the critical one, $T_c$. 
The contour line $T=1$, indicates the critical energy density,
$T_c$ where the phase transition (A) or the ignition (B) in the 
target is reached. This contour line is almost at a constant time,
indicating simultaneous transition or ignition in the whole target volume. 
If the target matter properties do not change significantly 
at this energy density, the heating continues as shown. If there is 
an ignition or melting, the dynamics may change drastically.
The two straight lines indicate the light-cones originating 
at the outer edges of the target at 
the moment when the irradiation
pulse ends. Note that the absorptivity in this case 
$\alpha_K (x) \not=\ \  $const. For more details please 
see Appendix B. 
}
\label{F-Integ}
\end{figure}

(ii) At the second step we add up the accumulated radiation 
at position $x$, to obtain the time dependence of 
energy density distribution, $T(t,x)$. This is done by integrating  
the deposited energy
$dU(t,x)$ from $t=0$, for each spatial position
\footnote{
Due to the flat target geometry and the light pulse strength degradation,
the irradiation of the central domains is about three times weaker
than in the previously demonstrated spherical configuration
\cite{CKP-LPB-18}.
}. 
 See Fig. \ref{F-Integ}.

Fig. \ref{F-Depos} shows the energy deposition per unit time
in the $t-x$ plane. 
Here the decrease of the irradiated energy due to the
absorption of the target is compensated by the increasing
nano-sphere density.
One can see that initially the
irradiation reaches only the near side of the target.
Later on the irradiation reaches
the target from both sides. Here the increased nano-shell density
ensures the uniform energy deposition in the middle
of the target.  

Fig. \ref{F-Integ} shows that the integrated energy deposition
is nearly homogeneous at time, $t \approx h/c$, when the 
deposited energy equals the critical value 
$T=T_c$. This contour  corresponds to a critical deposited energy
$Q$, which is sufficient for the ignition (B) or melting (A).

Thus, the increase of absorption in the middle layers is  
necessary to achieve a isochronous
burning front with time-like normal, i.e.
simultaneous volume ignition throughout the target.

First let us describe the space-time dynamics
for constant absorptivity
and without considering the energy loss due to the
light degradation during the penetration through the target.
(I.e. we assume that the beam energy current is much bigger
than the fraction deposited in the target. The
same assumption was used in ref.\cite{CKP-LPB-18}.)

Let us take the surface integral by using a delta function, selecting
the surface element, which can reach the given internal point at a time.
The deposited energy at space-time point at $(t,x)$ 
by a light front is $dU(t,x)\,dt\,dx\,S$ where $S$ is the transverse 
surface area. As the light front propagated 
with the speed of light the deposition length is $dx = c dt$, so
the space-time integral for the energy deposition is
\be
\int dU(t,x)\, \delta(ct \pm x) \,dt\,dx =
\int dU(t,x)\,dx,
\label{delta(tx)}
\ee
along the light front trajectory:
\be
(x - x_{l,r}) = \pm  (t - t_0) c, \ 
\label{traj}
\ee
where
$  x_{l,r} = \mp h/2 $, and it is assumed that the light front
starts from the left $(l)$ and right $(r)$ edge at time $t_0$.
Thus the deposited energy along the light beam trajectory can be 
characterized by the coordinate $x$ only. The deposition can be 
modified by changing absorptivity in the target by implanting nano-shells.

The average intensity of radiation reaching the surface of 
the DT pellet {\bf (B)} amounts to 
$dU$ per unit surface (mm$^2$) and unit time (mm/c).
Let us take a typical value for the total energy of the ignition
pulse to be  
$\sim$ 50 J, in time 1.04\, ps. 
Then  the radiation intensity from both sides is\
$dU \approx 
2 \cdot 50 {\rm J}\ \cdot 
0.0153^{-1} {\rm mm}^{-2}\ (0.526 {\rm ps})^{-1}$
 =  $  1.24  \cdot 10^{18}\ {\rm W / cm}^{2}$.

In case of the PLA target {\bf (A)} from both sides the radiation
intensity is 
$dU \approx 2 \cdot 15 {\rm mJ}\ \cdot 
0.81^{-1} {\rm mm}^{-2}\ (5.08 {\rm ps})^{-1}$  
$ =   7.28  \cdot 10^{11}\ {\rm W / cm}^{2}$.
The energy deposition dynamics is illustrated in Figs.
\ref{F-Depos} and \ref{F-Integ}.

At time $t$, the light can reach a space-time point ($t,x$), inside the 
flat target from the outside surface. At early times it may be that none 
of the spatial points at positive $t$, are
within the backward light-cone of a point ($t,x$). This domain is white in
Figs. \ref{F-Depos} and \ref{F-Integ}.
Before the time $h/$c there are points, which can be reached from at least
one side and there are others which are not. At time $h/$c all points of the 
flat target can be reached from both sides of the flat target. At this time 
the irradiation from the outside edges stops.
At late times, larger than $2h/$c,
the light pulses did reach all internal points ($t,x$) from 
both sides. At this time all energy deposition to the target is 
completed. See Fig. \ref{F-Depos}.
The choice of model parameters and their relations are described
in {\bf Appendix A}.

\section{Analytic Model}
\label{Analytic}

Let us take that the incoming irradiation energy at the flat target 
surface is $Q_0 = Q(h/2)=Q(-h/2)$.
In example {\bf B}, the total irradiation time is 
$t_x = 0.526$ ps, and  $Q_0 = 50$ J from one side and $ 2Q_0 = 100$ J 
from both sides together. We introduce the notation that the part of 
the energy deposited in the target is $D(x)$, at position $x$.
The remaining part of the beam energy, $Q(x)$, propagates further inwards
along the light beam trajectories (\ref{traj}).
The irradiation from the two sides is symmetric, and both decrease as
propagating inwards. Let us label the left and right sides by 
indices $l, r$. 

The sum of the deposited and propagating components of the radiation
satisfies the relations
$$
Q_r(x) + D_r(x) = Q_0 ,  \ \ \ \ \ 
Q_l(x) + D_l(x) = Q_0    .
$$
Note, that we neglect the energy reflected from the target surface. 
These propagating components change as 
\ba
dQ_l(x) = -\alpha_K Q_l(x)\,dx , \nnb
Q_l(x) = Q_0 - \int_{-h/2}^x \alpha_K Q_l(x') dx' \ , 
\label{de1}
\ea
where $\alpha_K$ is the absorption coefficient of the target.
In the case if $ \alpha_K$ is constant we get
\ba
Q_l(x)\!\! &=& Q_0 \ e^{-\alpha_K\,(x+h/2)}\ \ , \nnb
Q_r(x)\!\! &=& Q_0 \ e^{+\alpha_K\,(x-h/2)}\ \ . 
\ea
Here the second equation corresponds to the radiation from 
the surface at $+h/2$, propagating inwards in the $-x$ direction.

In the same case the deposited energies take the forms:
\ba
D_l(x)\!\! &= Q_0{-}Q_l =& Q_0 \left( 1 - e^{- \alpha_K (x+h/2)} \right) ,\nnb
D_r(x)\!\! &= Q_0{-}Q_r =& Q_0 \left( 1 - e^{+ \alpha_K (x-h/2)} \right) .
\ea
In the space-time domain where the point $(t,x)$ can be reached from both 
sides, the total deposited energy $D(x)=D_l(x)+D_r(x)$ is
\be
D(x) = 2 Q_0 \left( 1 - e^{-\alpha_K\cdot h/2}\ \cosh (\alpha_K\cdot x) \right).
\ee

Let us first
calculate the energy density, $dU_{l,r}(t,x)$, deposited at a 
space-time point ($t,x$), in the time interval $dt$ from earlier
times for the l.h.s. and r.h.s of the target. 
This is an integral of the quantity $D_{l,r}(x)$.
For a given point at $x$ we get the relation for the 
integrated energy density

\be
 U_r(t,x) =
 \int_{ah/(2c)}^{bh/(2c)} \left[ D_r(x) + D_l(x) \right]  d\tau  .
 \label{de2} 
\ee
The integral over $\tau$, runs from the nearest point of the backward 
light cone  $ah/(2c)$, 
to a later point, $bh/(2c)$, 
where the parameters $a$ and $b$ will be described below.  
Now let us introduce a dimensionless time variable:
$$
q \equiv 2\tau {\rm c}/h \ ,
$$
and with this notation, the boundaries, $a,b$, 
of the above integral are

\be
a = \left\{
\begin{array}{ll}
3/2{+}x/h, &	 q>3/2{+}x/h\\             
3/2{-}x/h, &    3/2{-}x/h<q<3/2{+}x/h \\   
1/2{+}x/h, & 	1/2{+}x/h<q<3/2{-}x/h \\   
1/2{-}x/h, & 	1/2{-}x/h<q<1/2{+}x/h \\   
0, 	   &	 q<1/2{-}x/h \\            
\end{array}
\right. \ .
\label{E4}
\ee
\be
b = \left\{
\begin{array}{ll}
3/2{+}x/h, &	q>3/2{+}x/h\\             
3/2{+}x/h, &    3/2{-}x/h<q<3/2{+}x/h \\   
3/2{-}x/h, & 	1/2{+}x/h<q<3/2{-}x/h \\   
1/2{+}x/h, & 	1/2{-}x/h<q<1/2{+}x/h \\   
1/2{-}x/h, &	 q<1/2{-}x/h \\            
\end{array}
\right. \ .
\label{E4}
\ee

In integrating over $dq$, we add the contributions
of those surface elements, from where radiation reaches the
internal point at $x$ at the same dimensionless time $q$. 
In the earliest case above the radiation
does not reach the point at $x$,  then in 
the second part the radiation 
from the closer side of the slab reaches 
$x$ but from the opposite point
not yet. In the third case radiation 
reaches $x$ from both sides, and so on.
In this way the space-time is divided 
into 4 domains. After the  irradiation
to the 4th domain is completed the 
warming up of the target stops because 
the irradiation finishes in a limited time period.

Thus the integrated energy deposited up to the 
dimensionless time $q$ on the r.h.s. 
of the system (i.e. for  $x>0$ ) is
\ba
&&U_r(x,q) = 
\nnb
 &&\!\!\!\!\left\{
\begin{array}{ll}
2 Q_0 \left( 1 - e^{-\alpha_K\cdot h/2}\ 
\cosh (\alpha_K\cdot x) \right) \ ,\\	
&\hskip -4cm   q>3/2{+}x/h \\[3mm]      
Q_0 \left(1-e^{+\alpha_K (x-h/2)} \right) 2x/h\  +&\\
2 Q_0 \left(1-e^{-\alpha_K\, h/2}  
\cosh(\alpha_K x)\right)\\ 
(1{-}2x/h)  \ + &\\
Q_0 \left(1-e^{-\alpha_K (x+h/2)} \right) 
(q-3/2+x/h)\ , &  \\
&\hskip -4cm  3/2{-}x/h<q<3/2{+}x/h \\[3mm]      
Q_0 \left(1-e^{+\alpha_K (x-h/2)} \right) 2x/h\ +&\\
2 Q_0 \left(1-e^{-\alpha_K\, h/2}  
\cosh(\alpha_K x)\right) \\ 
(q{-}1/2{-}x/h) \ , &\\
& \hskip -4cm 1/2{+}x/h<q<3/2{-}x/h \\[3mm]      
Q_0 \left(1-e^{+\alpha_K (x-h/2)} \right)  
(q-1/2+x/h)\ , &  \\
&\hskip -4cm  1/2{-}x/h<q<1/2{+}x/h \\[3mm]      
	0 \ , & \hskip -4cm q<1/2{-}x/h \        
\end{array}
\right. \nnb
&&\!\!\!\! {\rm for}\  x>0 \ ,
\ea
and $U_l(x,q) = U_r(-x,q) $ for $x\le 0$.
\smallskip

Neglecting the compression and assuming 
constant specific heat $c_v$,
we get that $k_B\,dT = \frac{1}{n\,c_v} dU$,
where $k_B$ is the 
Boltzmann constant, and $n$ is the number density of the 
target matter. Therefore
\ba
&& k_B\, T(t,x) = \frac{1}{n\,c_v} \cdot U(x,q)  = \sigma  \times
\nnb
 &&\!\!\!\!\left\{
 \begin{array}{ll}
 2  \left( 1 - e^{-\alpha_K\cdot h/2}\ 
\cosh (\alpha_K\cdot x) \right) \ ,\\	
&\hskip -5cm   ct>3h/4{+}x/2 \\[3mm]      
 \left(1-e^{+\alpha_K (x-h/2)} \right) 2x/h\  +&\\
2  \left(1-e^{-\alpha_K\, h/2}  
\cosh(\alpha_K x)\right)\ 
(1{-}2x/h)  \ + &\\
 \left(1-e^{-\alpha_K (x+h/2)} \right) 
(ct/h-3/2+x/h)\ , &  \\
&\hskip -5cm  3h/4{-}x/2<ct<3h/4{+}x/2 \\[3mm]      
 \left(1-e^{+\alpha_K (x-h/2)} \right) 2x/h\ +&\\
2  \left(1-e^{-\alpha_K\, h/2}  
\cosh(\alpha_K x)\right)\\ 
(ct/h{-}1/2{-}x/h) \ , &\\
& \hskip -5cm h/4{+}x/2<ct<3h/4{-}x/2 \\[3mm]      
 \left(1-e^{+\alpha_K (x-h/2)} \right) 
(ct/h-1/2+x/h)\ , &  \\
&\hskip -5cm  h/4{-}x/2<ct<h/4{+}x/2 \\[3mm]      
0 \ , & \hskip -5cm ct<h/4{-}x/2 \ ,        
 \end{array}
 \right. \nnb
\ea
where $\sigma$ is defined as 
\be
\sigma \equiv \frac{Q_0}{n\,  c_V}\, . 
\ee
After substitution of 
the number density of uncompressed DT ice
$n= 3.045 \cdot 10^{22}$ cm$^{-3}$, we get
$
\sigma= 16.42 \cdot 10^{-22} {\rm J/cm^3}.
$

In case of constant $\alpha_K$ one can achieve time-like detonation
only in the space-time domain which receives irradiation from both 
sides, i.e. for $ct$ values, which satisfy the relation
$h/4{+}x/2<ct<3h/4{-}x/2$  (in the r.h.s. 
space-time domain). One can show that in this case
\ba
&k_B \partial T(t,x)/\partial x  =  \nnb
& e^{-\alpha_K\,h/2} \frac{2\sigma}{h}
\left[\phantom{\frac{1}{1_1}}\!\!\!\!\! (1-x/h)e^{\alpha_K\,x}+ 
\alpha_K \sinh(\alpha_K x)\right.\nnb
&\left. 
\left(\frac{h}{2}+x+\frac{1}{\alpha_K}\coth(\alpha_Kx)-ct \right)
\right] 
\ea
\be
k_B \partial T(t,x)/\partial (ct)   = 
\frac{2\sigma}{h} \left(1-e^{-\alpha_K\, h/2} \right) \cosh(\alpha_K x) \ .
\ee

The surface of the ignition is characterized by 
the energy density, $T(ct,x) = T_c$, contour line, where 
$T_c$ is the ignition energy density.
The tangent of the $T(ct,x) = T_c$ contour line, for the points 
separating its space-like and time-like 
parts  is
\be
 {\left( \frac{\partial x}{\partial ct} \right)}_{ T_c} = 
 {\left( \frac{\partial T}{\partial ct} \right)}_{ T_c} \Bigg/ 
 {\left( \frac{\partial T}{\partial x} \right)}_{ T_c} = \pm 1 \ .
\ee
So the point $(t_c,r_c)$ where the space-like and time-like 
parts of the surface meet (for 
${\left( {\partial x}/{(c\partial t)} \right)}_{T_c} = + 1$)
is :
\be
c t_c =  \zeta +\frac{h}{2} +x_c 
\ee
where
\ba
&\zeta=\left[\left(1{-}x/h\right)e^{\alpha_K\,x}+\cosh(\alpha_K\,x_c)+\right.\nnb
&\left. \left(1-e^{\alpha_K\, h/2} \right) \cosh(\alpha_K x_c)\right] \times \nnb
&\left [ \alpha_K \sinh(\alpha_K x_c) \right]^{-1}\ .
\ea

The ignition starts at $x=\pm h/2$ and it propagates 
first slowly inwards. Due to the radiative heat transfer 
the contour line of ignition, $T(t,x) = T_c$,
accelerates inwards, and at $x_c = x_c (T_c)$ it develops smoothly 
from space-like into a time-like ignition hypersurface.

It is interesting that
similar type of gradual development from 
space-like into time-like 
regime  occurs in the {\em hadronization} 
phase of ultra-relativistic heavy ion collisions
\cite{FO-HI,HI-2,HI-3,HI-4,HI-5,HI-6,HI-7,HI-8}.
If we include radiative heat transfer in our model, 
the transition from space-like 
to time-like behavior will be gradual. This, however, 
requires more involved numerical calculations.

The hyper-surface, where the energy density,
$T(t,x)=$const. obtained from the analytical solution
shows that time-like behaviour occurs only in the central domain of the
flat target. Therefore we cannot achieve uniform volume ignition, and 
instabilities might develop!

\section{Variable absorptivity}

In order to study the effect of variable absorptivity 
we reformulated 
the numerical model to perform all integrals of 
the model numerically.
This enables us to study  
pellets with nano-shells inside.

In ref. \cite{Hu2014}, the Rochester and NIF experimental data were 
analysed by using opacity data, extracted from basic principles   
and from comparison with ICF experiments. The absorption coefficient
$\alpha_k$ (cm$^{-1}$), defined by $I(x) = \exp(-\alpha_k\,x) I_0$, 
and the Rosseland and Planck opacities, $K_r$ defined
by  $I(x) = \exp(-\alpha_{K_r}\,\rho\,x) I_0$, were estimated and used
to simulate  ICF direct ignition experiments.

In our previous calculations we used the absorption coefficient
$\alpha_k$ (cm$^{-1}$), and the approximation that the intensity of the
incoming laser light flux is sufficiently large, so that its decrease by the
absorption is negligible \cite{CKP-LPB-18}. Here we have relaxed this 
constraint. Also in the previous section we assumed a constant absorptivity,
 $\alpha_K = $const. Relaxing this simplification changes the situation
essentially as it was shown \cite{CKP-LPB-18}.

Our equations for $Q_l(x), Q_r(x), D_l(x), D_r(x)$ and $D(x)$, should now
be evaluated numerically based on the same differential equations that 
were used for the constant absorptivity case.

With increased absorptivity one could reach more rapid heating or 
energy deposition. 
Substituting the fusion cross section, the reactivity is increasing
up to an energy density analogous to about $T = 70-100$ keV temperature, 
and then decreases again. Thus, we could
aim for a heating up to this energy density with increased light 
absorption, with smaller pre-compression.

\section{Absorptivity by Nanotechnology} 

Doping ICF pellets with golden nano-shells 
enables us to achieve the desired variable absorptivity
\cite{Tana2016}.
Nano-shells irradiated by laser light exhibit a resonant
light absorption, which can increase the plasmon field-
strength by up to a factor of 40-100 or more
\cite{N_Halas,N-2}.
At present
experimentally realizable nano-shell sizes range for core
sizes of 5-500 nm, and for shell thickness of 1-100 nm.

The resonant light frequency of the nano-shell can be
tuned in a very wide range by changing the 
size $(r_{outer})$  and thickness ratio of the nano-shell. 
If the core $(r_{inner})$ versus the shell thickness, 
$r_{inner}/(r_{outer}-r_{inner})$, is changed from 2 to 800 the resonant
wavelength changed from $0.5$ to $ 10^4$ nm \cite{Loo04}
with radiation by a beam of specific polarization. 

In case of flat targets with irradiation in one direction only,
we can use resonant nano-rods instead of spherical nano-shells.
In this case, due to the transverse polarization of the light beams,
the resonant nano antennas should lay in the transverse plane. 

At the resonant frequencies the
nano-shells are able to absorb resonantly a rather high
portion of incoming light. We can introduce the $absorption$,
$scattering$ and $extinction$ efficiencies, 
$Q_{abs}$, $Q_{sca}$,  and $Q_{ext}$,
respectively \cite{Lee05}. These coefficients, $Q_i$, describe how
much part of the energy of the incoming light is 
absorbed or scattered by the nano-shell, compared to its
geometrical cross-section, G, i.e. for a sphere of radius
$R$, $G = R^2\, \pi$.

The nano-shells can be tuned to achieve either larger absorption
efficiency or larger scattering efficiency. For our purposes
the larger absorption efficiency is more important. Note that the resonance
extinction or absorption efficiency can reach a factor
10 or even more \cite{Alam06}.

\begin{figure}[h]  
\resizebox{1.0\columnwidth}{!}
{\includegraphics{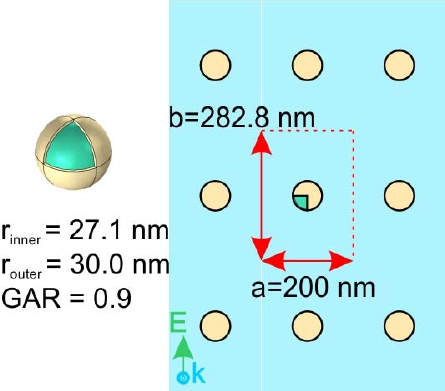}}
\caption{
(color online) 
Left: Schematics view of a single core-shell nano-sphere. 
Right: Rectangular lattice of such nano-spheres in a transverse 
layer of the target.
The beam directed spatial density distribution of monolayers
is taken to be Gaussian. 
}
\label{pfa4}
\end{figure} 

Most simple materials at room temperature have
small absorptance, $\alpha_0 \le 1 $cm$^{-1}$
at and around visible light frequencies.
The only exceptions are light frequencies that
are resonant to some molecular excitation
frequencies of the given material.
Dense and hot plasma has also similarly
low absorptance at high light frequencies
above $\hbar\omega \ge 10$ keV. Thus, the
initial low temperature target with moderate
pre-compression is sufficiently transparent
for laser light. This then, with well distributed
nano-shells compresses and increases the energy
density of the plasma rapidly and simultaneously
to ignition level. By then, the hot and dense
plasma becomes transparent to radiation, which is
still not isotropic but beam directed, just like
the accelerated target ions. These latter ones
collide with each other in a colliding beam
configuration. 

The absorptivity of the target can be
regulated by the density of implanted nano-shells. Let us denote that the
absorption coefficient of the DT fuel target by $\alpha_{k0}$ (cm$^{-1}$).

If we implant nano-shells, then the absorptivity will increase to
\be
\alpha_{k} = \alpha_{k0} + \alpha_{ns}\ . 
\ee
Here the absorptivity of nano-shells is 
\be 
\alpha_{ns} = \rho\ G\ Q_{abs} \ ,
\ee
where $G=R^2 \pi$, and the nano-shell density is $\rho$ (cm$^{-3}$).

\begin{figure}[h]  
\resizebox{1.0\columnwidth}{!}
{\includegraphics{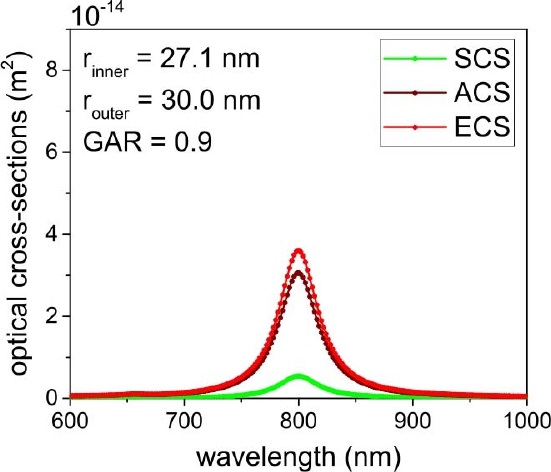}}
\caption{
(color online) 
Optical cross-section of an individual core-shell nano-sphere
optimized to absorb light at 800 nm wavelength and 
optical response of the same core-shell nano-spheres composing a
rectangular lattice. The Scattering Cross Section (SCS), 
Absorption Cross Section (ACS) and Extinction Cross Section
are shown.
}
\label{pfb5}
\end{figure}

For nano-shells of $R=30$ nm and wall thickness of 3 nm
$G=2.83\ 10^{-11}$ cm$^2$, we obtained from electromagnetic field
calculations  the absorption cross section, see Fig. \ref{pfb5}.
For a typical nano-shell density 
\cite{James07} of 
$\rho = 10^{11}/$ cm$^3$ 
we obtain an additional absorptivity due to nano shells
\cite{Csetal} of
\be 
\alpha_{ns} = \rho G Q_{abs} =  115 \ {\rm cm}^{-1}\ .
\ee

Based on our computations, we found that 10 layers of sparse core-shell 
nano-sphere lattices are capable of resulting in a total absorption of 
33\% and 77\% in case of uniform and predefined Gaussian distribution. 
Even larger absorption improvement can be achieved up to 98\% by 
optimizing the position of the layers. The absorption can be further 
improved to 66\% and 99\% via double sided illumination of layers 
with uniform and optimized Gaussian distribution, respectively. As a 
result the total absorptivity of the target can be enhanced up to 
115 cm$^{-1}$ / 139 cm$^{-1}$ \& 
217 cm$^{-1}$ / 216 cm$^{-1}$ \& 
216 cm$^{-1}$ / 291 cm$^{-1}$ 
via single / double sided illumination of layers with
uniform - predefined Gaussian - optimized Gaussian distribution. 

On the other hand, we have found
(Appendix B), that for optimal simultaneous 
ignition we need only an additional maximum absorptance
from nano shells
\be 
\alpha^C_{ns} + \alpha_{ns}(0) = 11.6\ {\rm cm}^{-1} ,
\ee
in the center of the target. For this central absorptance,
assuming that the absorptivity is linearly proportional 
with the nano-sphere density, we need a central density of
of nano-shell density of
$\rho = 1.03\ 10^{10}/$ cm$^3$, 
and these nano-spheres have a volume of 
$V_{ns} = 1.16\ 10^{-6} $ cm$^3$ in 1 cm$^3$.
Of course, with
higher nano-shell density, higher absorption efficiency can also
be achieved. In case of linear configuration, resonant 
nano-rods, which are parallel 
to the direction of E-field of polarized irradiation, may
provide even larger amplification of the light 
absorption. 

We have studied how the target absorptivity is 
improved via core-shell type plasmonic
nano-shells, via solving the Maxwell equations,
and evaluating the ohmic heating. A Deuterium target
was considered  with periodic boundary 
conditions in the transverse directions, 
and we took the target thickness as 
$h=$ 0.11 mm. 

Previous studies in the literature have 
shown that core-shell particles of the same
composition are capable of resulting in resonance 
at a specific wavelength in case of different
generalized aspect ratios (GAR).

The thin (thick) shell 
composition results in a narrow (wide) effective scattering
cross-section, which is accompanied by large (small) 
absorption \cite{tam07}. As mentioned above, the thin shell 
composition increases the absorptivity. 
Accordingly, the generalized aspect ratio,
$GAR=r_{inner}/(r_{outer}-r_{inner})$, 
of silica-gold core-shell particles has been tuned to 
$GAR=0.9$, in order to get peak  absorption
and scattering simultaneously at 
the 800 nm central wavelength of the 1 ps laser
pulse. In our calculation the nano-sphere particles 
are arranged in a sparse, rectangular lattice, so that 
the near-field interactions are  weak,
while the lattice resonances do not affect the 
spectra around 800 nm for this strongly
sub-wavelength period (Fig. \ref{pfa4}). As a result, a 
mono-layer exhibits an absorptivity peak at the
central wavelength (Fig. \ref{pfb5}).
Based on our computations, we found that
10 layers of sparse core-shell nano-sphere lattices
are capable of resulting in a total absorption
larger than 80\% of the incoming laser pulse. However, 
the absorptivity can be
further improved to 85\% via double sided, compared to 
single sided irradiation by tuning the
phase of the counter propagating waves. 
Even larger absorptivity improvement can
be achieved up to 99\% by optimizing the position 
of the layers. As a result, 
the total absorption can approach 98\% and 99\%, and the target
absorptivity can be enhanced up to 100 cm$^{-1}$ and 
250 cm$^{-1}$ via phase tuning and position
optimization, respectively.
A detailed report of these studies will be published
separately \cite{Csetal}.

For DT targets, with densities in the range of 
$\rho = 5 - 200$ g/cm$^3$, and $T \approx 10^5$ K, absorptivities 
are obtained in the range of 
$$
\alpha_{k0} = 10^{-1} - 10^2 \ {\rm cm} ^{-1} ,
$$ 
when the light frequency increased to  $\hbar\omega = 1-10$ keV
\cite{Hu2014}.  
The typical light mean free path is about
1-10$^4\ \mu$m. Thus, while for low frequency radiation, 
$\hbar\omega = 1 - 100$ eV 
the hot and dense DT target is quite opaque, at higher frequencies or energy,
it is much more transparent. This leads to the result that the
initial lower energy pulse leads primarily to
compression. This effect could be enhanced further with the application of
the thin ablator sheet on the surface of the pellet
\cite{Benredjem2014}.

The additional opacity of nano-shells with typical nano-shell densities
can increase the absorptivity up to the values
$$
\alpha_{ns} = 10 - 10^2 \ \ {\rm cm}^{-1} \ .
$$
This makes the fast ignition possibilities very versatile in
this light frequency range. 
We can experiment with  variable absorptivity, which is the 
normal absorptivity of the DT fuel, 
$\alpha_{k0} \approx  1$ cm$^{-1}$ and an additional constant 
absorptivity from the nano-shells
$\alpha^C_{ns} = 3.2$ cm$^{-1}$
at the front and back edge of the pellet 
(i.e. at $\pm h/2$) while in the center,
$\alpha_{k0} + \alpha^C_{ns} + \alpha_{ns}(0) =
\ \ $ 30.3 cm$^{-1}$. 

The space time profile of the ignition, will then depend
on the profile of the nano-shell doping towards the
center of the pellet. See Fig. \ref{F-Integ}.

One can optimize this profile to achieve the largest
simultaneous volume ignition domain, which eliminates the 
possible development of instabilities.

Apart of the large cross section of plasmon excitations, 
which can enhance the geometrical cross section of
the nano-antennas by several orders of magnitude, surface 
plasmon resonance excitation is a much more efficient process for
the quantum mechanical "hot carrier" generation. This process
is much faster than the heating via the external hohlraum, and its
typical time-scale is around 10-500 fs \cite{Liu-ea-2018}. This is by
orders of magnitude shorter than the irradiation pulse length.
Furthermore, at high irradiation intensities, when the temporal 
spacing between incident photons becomes smaller than the hot carrier
relaxation time, the photoluminescence blue-shifts the radiation frequency
at large light intensities, due to a multi-photon quantum effects.
These two additional effects make the heating with nano-antennas
very fast and more effective. These effects
depend on the features of the implanted nano-antennas, which can
be calculated quantitatively \cite{Csetal}.

Planar metal surfaces reflect most of the incident light, 
and in this case light absorption is not very efficient. Nevertheless,
light absorption can be further enhanced by exciting localized 
surface plasmon resonances. This produces an antenna 
effect resulting in light collection from an area that is 
larger than its physical size.
Surface nano-wire arrays may additionally reduce this reflection 
and can give rise to higher energy densities with a 0.5J laser pulse
\cite{Purvis2013,Nano-Rods}.

Here the question arises how the doped nano-shells will 
behave under compression. The optimal thickness 
and radius of the golden nano-shells are chosen to achieve resonance 
condition of the laser irradiation. The stiffness of the shell interior,
can be chosen to keep the spherical shape of the nano-particles.
From studies of ICF capsules
employing DT wetted foam \cite{Sacks1987} of similar size small pores, 
concluded that the in-flight aspect ratio, hydrodynamic efficiency, 
and hot electron tolerance can be at least as good as for conventional
DT targets. The compressed nano-shells will be resonant to higher
laser frequencies, which can be taken into account during the nano-shell
production.

In the case of nano-rods implanted in the transverse plane with 
respect to the direction of the light beam, the nano-antenna 
parameters are not effected essentially by the compression. 
This makes the tuning of the nano-antennas simpler.

\section{Progress of ignition}

Depending on the density at ignition,
the rate of fusion reactions can be approximated as
$
n_D n_T   < \sigma \vec{v} >,
$
which may lead to a full DT burning time
after the moment of ignition, 
of the order of 10 ns. This time
is 3 orders of magnitude longer than the time to reach uniform ignition
with the proposed time-like ignition method. During this time interval 
the expansion 
of the hot burning plasma can be substantial, which leads to
reduced burning rate and a highly reduced burn-up fraction.

Several physical processes may remedy this problem. The first is 
pre-compression, as mentioned in the introduction. 
An initial compression 
in the beam direction by a factor of 10, increases the fusion rate
by a factor of 100, while the laser irradiation pulse with the same
pulse energy should be shortened to 53 fs. The beam intensity should
also be increased by an order of magnitude to 
$dU \approx 1.24 \cdot 10^{19}\ {\rm W / cm}^{2}$.

This feature, the ion pre-compression by the LWFA configuration.
was tested in recent experiments in linear configuration, and
reached substantial compression comparable to the one reached at NIF
\ \cite{Bonasera2019}.
%
\footnote{For comparison the 3D NIF
experiments use a linear size
compression of ~30 and a density increase of
almost 1000.}

For more realistic estimates
we need relativistic fluid dynamical, PICR analysis of the dynamics of the
compression and expansion of the system. Possible 
pre-compression 
can be achieved by an weaker and longer (so called "low foot") initial 
irradiation, but still without an ablator layer. The final expansion
after the onset of the burning phase, can be suppressed by a continued
final "low foot" radiation (in the beam direction). The energetic 
short ignition pulse accelerates the target matter inwards also, as
the deposited light quanta provide an inward momentum. These effects
can be evaluated by PICR, relativistic fluid dynamics or
relativistic molecular dynamics calculations.

The microscopic features of the ignition increase the chances
for faster burning also. The initial motion of the ions at the start
of the burning is beam directed and not thermal. 
It is also enhanced by the laser wake-field
acceleration of the ions. Furthermore, non-thermal 
laser driven plasma-blocks
\cite{Hora2017,Sau1996}
due to ponderomotive EM forces may additionally accelerate
ions in the beam direction to relativistic velocities. This significantly
increases the fusion rate, particularly in case of two opposite beams. 
These effects are recently experimentally verified \cite{Bonasera2019}.

For optimal choice of a fusion method such effects will be modeled
and optimized before experimental tests.

\section{Conclusions and discussions}

Using nano-technology for ICF was recently discussed
\cite{Nano-Rods,Nano-2}.  Placing aligned nano-rods or nano-wires
on the surface of the pellet and irradiating it with 
femtosecond laser pulses of relativistic intensity, leads to a 
plasma with large electron intensity and pressure. However, this
pressure would lead to a pressure driven adiabatic compression and
heating, which can lead to Rayleigh-Taylor instabilities, preventing
simultaneous volume ignition.

In our model calculations, we have neglected compression as well as 
the reflectivity of the target matter.      
The relatively small absorptivity made it possible that the radiation
could penetrate the whole target.  
Our estimated 
ignition energy is 208 keV/mg, 
corresponding to the contour $T=1$ in Fig. \ref{F-Integ}. 
We see that the critical ignition energy density is 
{\bf reached in about 80\% of the target volume simultaneously}
(i.e. on a time-like hyper-surface).
 In this domain no instabilities may occur.

We can also apply this model to a moderately
pre-compressed target, which is transparent and has
larger absorptivity. In this situation the ignition energy density can be
somewhat smaller, but we still can optimize the pulse strength
and pulse length to achieve the fast, nearly complete ignition of the 
target.
We have considered relatively short and very intensive irradiation, which 
is relevant for achieving simultaneous, time-like ignition (B). 
This method leads to a faster energy density
increase towards $T_c$ than the growth rate of 
the Rayleigh-Taylor instability or
other instabilities. This can make even an order of magnitude
increase in the critical irradiation time. Furthermore if we 
also consider the smoothing effect of radiation dominated ignition
front (B) then the critical irradiation time can be even longer. This
could only be estimated with 3+1D RFD calculation.

Finally we want to point out that
if we neglect the relativistic effects, 
the theory would be far-fetched from reality. 
It is important to use the proper relativistic treatment
to optimize the fastest, more complete ignition, 
with the least possibility of instabilities, which reduce 
the efficiency of ignition.

\section*{Acknowledgements}

Enlightening discussions with 
Istv\'an F\"oldes and Gerg{\H o} Pokol, P\'eter R\'acz, 
Kirill Taradiy and S\'andor Varr\'o, 
are gratefully acknowledged.
This work is supported in part by the Institute of Advance Studies, 
K{\H o}szeg, Hungary, and the Frankfurt Institute for Advanced Studies.

\section*{Appendix A: Parameters}

Based on the NIF results
\cite{NIF2018}
the necessary ignition energy of the DT target {\bf (B)} is 
$ Q/m = 207.7$ kJ/mg.
Then assuming a 
$ Q_0 = 100$ J 
laser pulse energy, we can ignite a DT target of mass
$m = 0.481$ $\mu$g.\
The density of DT ice is 
$ \rho = 0.225$ g/cm$^3$,
which leads to a target volume of
$V = 0.00214$ mm$^3$.
For a minimal target surface, the diameter and
height of the DT target cylinder becomes
$2R = h = 0.111$ mm, 
and its cross section is
$A = 0.0153 $ mm$^2$.  

The critical energy density for ignition is:\\
$\epsilon = \rho \cdot Q/m  = 46.47$ MJ/cm$^3$ (kJ/mm$^3$),\\
while the required pulse duration is \\
$t_{pulse} = h/c_{DT} =  0.526$ ps. 
During this time interval we should deposit the 
total pulse energy, $Q_0$ to the target, which leads to an initial 
energy flux, at the flat surface of the target 
$u_0 = \epsilon/t_{pulse} = \epsilon c_{DT} /h = 88.8$ kJc/mm$^4$.
Therefore we get $U_0 = A u_0 = 1.359$ kJc/mm$^2$.

%
\begin{figure}[h]  
\resizebox{1.0\columnwidth}{!}
{\includegraphics{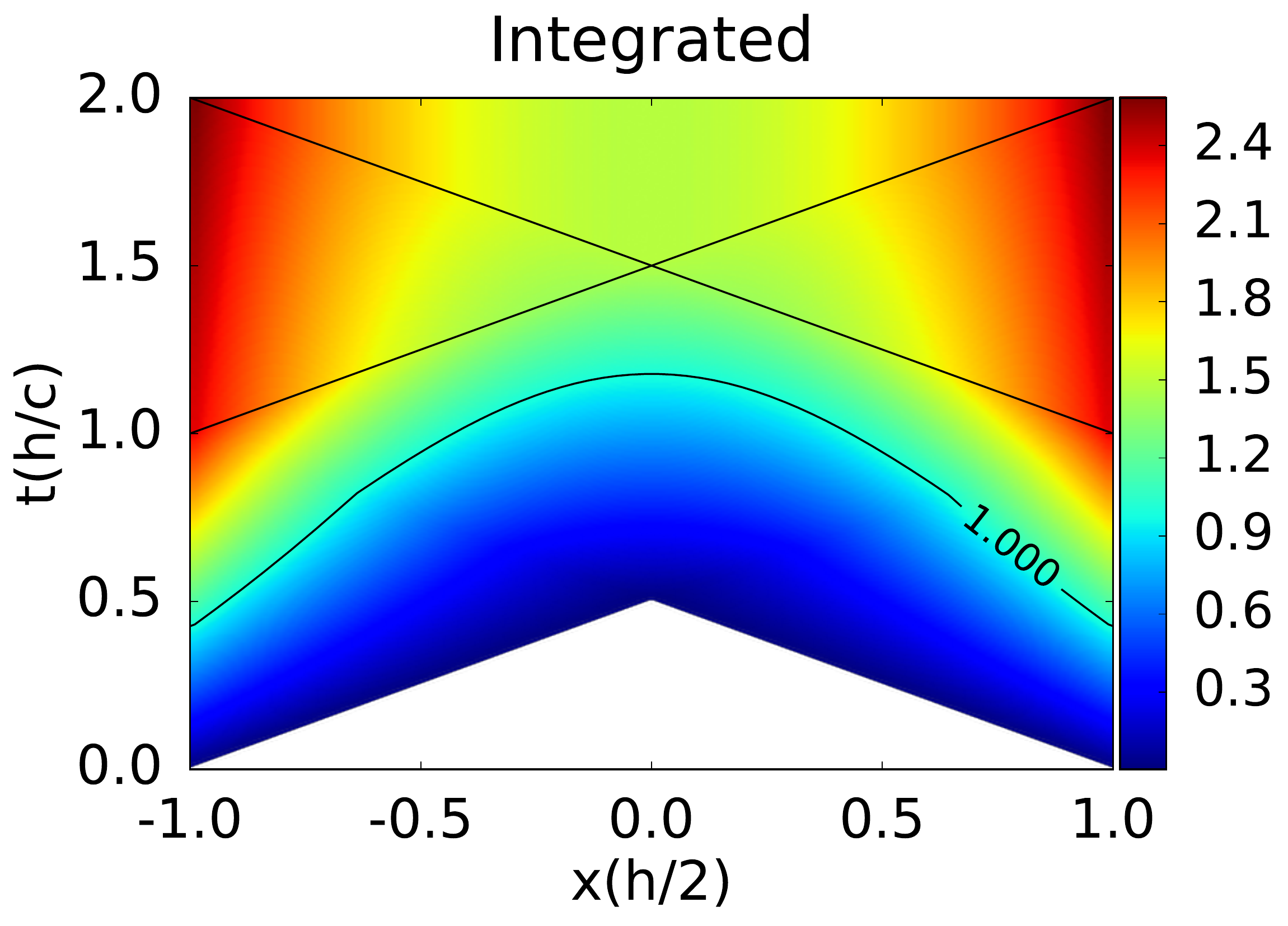}}
\caption{
(color online)
Same as figure \ref{F-Integ}, but without nano-shells.
The integrated energy up to a given time in the space-time
across the depth, $h$, of the flat target. The color strip
indicates the energy density, in units of the critical 
energy density $(T_c)$. 
The contour line $T=1$, indicates points 
where the phase transition or the ignition in the 
target is reached. 
This contour line, compared to the one in Fig. \ref{F-Integ}, is 
never constant in time, indicating no simultaneous whole 
volume transition or ignition. The time-like (causally unconnected)
part of the transition takes place only in the central $\sim$ 15\% of the 
target volume.
The two straight lines indicate the light-cones originating 
at the outside edges of the target at the ending of the irradiation
pulse.
}
\label{F-Integ-const-alpha}
\end{figure} 

Taking into account that in the beam direction the light front
propagates with the speed of light (in the target material), we get 
the deposited energy
\begin{equation*}
\int u(x,t) d^3x dt = \int Au(x,t) \delta(x-ct) dx dt,
\end{equation*}
and we can introduce the linear deposited surface energy density,
$ A u(x,t) $, inside the target  as
\begin{equation*}
	D(x) dx = \int_{t_0} A u(x,t) \delta(x-ct)\, dt\, dx\,  ,
\end{equation*}
along a light beam starting at $t_0$.
From the incoming surface energy density $ Q(x) $ 
a part, $\alpha_K(x)$, is deposited in the target material:
\begin{equation*}
	D(x) dx =  \alpha_K(x) Q(x) dx \ ,
\end{equation*}
and the remaining lesser part continues to propagate along
the light beam as described in Section \ref{Analytic}.

\subsection*{Appendix B: Absorptivity}

For constant absorptivity $\alpha_K$ 
only slow heating of the target is possible, 
see Fig. \ref{F-Integ-const-alpha}.

In our numerical calculations with nano-shells implanted 
in the target to increase central absorption (see 
Figs. \ref{F-Depos} and \ref{F-Integ}), we used the distribution:
\begin{equation*}
	\alpha_{ns}(s) = 
\alpha^C_{ns} +
\alpha_{ns}(0) \cdot \exp \left[4\times 
\frac{ 
\left(\frac{s}{100} \right)^2}
{\left( \frac{s}{100}-1\right) \left( \frac{s}{100}+1\right)} 
\right] \, .
\end{equation*}
Here the length scale, $s$, is chosen so that 
$s=100$ corresponds to 
$x = h/2 = 0.0555$ mm.
The edge absorptivity of the fuel is 
$\alpha_{k0} = \ \ $ 1.0 cm$^{-1}$, then
from the nano-shells
$\alpha^C_{ns} = \ \ $ 9.10 cm$^{-1}$, and
$\alpha_{ns}(0) = \ \ $ 20.2 cm$^{-1}$.
Thus in the centre the absorbance is   
$\alpha_{k0} + \alpha^C_{ns} + \alpha_{ns}(0) =
\ \ $ 30.3 cm$^{-1}$. 
For this absorbtance parameters we need a modest density of
nano-spheses embedded into the target fuel. In the center 
$
\rho_{ns}(0) = 2.48\ 10^{10}\ $cm$^{-3}\ ,
$
while at the outside edges even less
$
\rho_{ns}^E = 8.05\ 10^{9}\ $cm$^{-3}\ .
$
In the center of the target material the nano-shells 
will occupy only 
$ 2.9\ 10^{-4}$ \%
of the volume.

With these absorption parameters only 0.25\% of the energy
of the incoming laser pulse reaches the opposite side of the 
target. Thus, the expectation is an energy balance with
a possible minimum of loss.

\bigskip

\section*{References}



\end{document}